\documentclass[prb,aps,twocolumn,showpacs]{revtex4}
\usepackage{epsfig}
\begin{document}

\title{Electron Removal Self Energy and Its Application to Ca$_2$CuO$_2$Cl$_2$}

\author{Chul Kim$^1$, S. R. Park$^1$, C. S. Leem$^1$, D. J. Song$^1$,
H. U. Jin$^1$, H.-D. Kim$^2$, F. Ronning$^3$ and C. Kim$^{1,*}$}

\affiliation{$^1$Institute of Physics and Applied Physics, Yonsei
University, Seoul 120-749, Korea}

\affiliation{$^2$Beamline Research Division, Pohang Accelerator
Laboratory, POSTECH, Pohang, Kyungbuk 790-784, Korea}

\affiliation{$^3$MPA-10 Division, Los Alamos National Laboratory,
Los Alamos, New Mexico 87545}

\date{\today}

\begin{abstract}
We propose using the self energy defined for the electron removal
Green's function. Starting from the electron removal Green's
function, we obtained expressions for the removal self energy
$\Sigma^{ER}$(\textbf{k},$\omega$) that are applicable for
non-quasiparticle photoemission spectral functions from a single
band system. Our method does not assume momentum independence and
produces the self energy in the full \textbf{k}-$\omega$ space.
The method is applied to the angle resolved photoemission from
Ca$_2$CuO$_2$Cl$_2$ and the result is found to be compatible with
the self energy value from the peak width of sharp features. The
self energy is found to be only weakly \textbf{k}-dependent. In
addition, the Im$\Sigma$ shows a maximum at around 1 eV where the
high energy kink is located. 
\end{abstract}
\maketitle

\section{Introduction}
Photoemission lineshape contains information on the many-body
interactions in the solid under study. One notable example is the
kink structure in the angle resolved photoemission(ARPES) spectral
function from a metal at the bosonic mode energy due to
electron-bosonic mode coupling\cite{McDougall,Hengsberger,Valla}.
In this regard, the kink structures at near 70 meV found in the
ARPES spectral functions from cuprate high temperature
superconductors (HTSCs) have been a hot topic during the last
several years as it may provide a clue on what mediates the
pairing in the
HTSC\cite{Bogdanov,Lanzara,Kaminski,Gromko,Meevasana}. It is still
under debate whether the kink is caused by phonons or magnetic
excitations. In addition to the kink near 70meV, an anomalous high
energy kink structure from insulating cuprate Ca$_2$CuO$_2$Cl$_2$
(CCOC) was recently reported\cite{Ronning}. Subsequently, it was
found that the high energy kink also exists in doped cases, making
it omnipresent at all dopings\cite{Xie,Graf}. Existence of the
high energy kink reveals that there may be another energy scale
from the many-body interactions in HTSCs. It is thus important to
find the physics in this high energy scale.

One can extract from the data the information on the many-body
interactions through the self energy analysis. Conventionally, the
self energy analysis is done by extracting the dispersion and peak
width from the momentum distribution curves
(MDCs)\cite{McDougall}. However, there are underlying assumptions
for the analysis to be valid. First of all, ARPES peaks should be
sharp enough to be quasiparticle-like. Secondly, the bare band
dispersion should be quite linear within the energy range of
interest so that one can \textquotedblleft guess\textquotedblright
the bare band dispersion. Unfortunately, none of these are valid
assumptions in the analysis of ARPES spectral function from CCOC:
The peaks are too broad to be considered as quasi-particle
peaks\cite{C.Kim,K.M.Shen} and there is an inherent confusion
regarding what should be considered as the original bare band as
CCOC is a Mott insulator. Similar limitations also apply to the
recently found high energy kinks in HTSCs\cite{Xie,Graf}. However,
the lack of quasiparticles should not be confused with a lack of
physics. For example, in Luttinger liquids, where Fermi liquid
quasiparticles do not exist, ARPES has provided detailed
information on the excitation spectrum.\cite{B.J.Kim}.

It is therefore necessary to develop a method to extract the self
energy from non-quasi particle ARPES spectral functions. In fact,
there was an earlier attempt to obtain the self energy through a
Hilbert transformation with the assumption of particle-hole
symmetry\cite{Norman}. This method however is not applicable to
systems without particle-hole symmetry (for example,
CCOC\cite{Hasan}). Here we propose to use the self energy defined
for the electron \emph{removal} Green's function. Even though
different, it bares similar characters of the usual self energy
and its imaginary part is found to appropriately represent the
width of the spectral feature. Moreover, we developed a method to
extract the self energy from a general ARPES spectrum from a
single band system. It can thus be used to for spectral functions
with broad spectral features. We believe our analysis method
provides a new way to examine the underlying physics from
photoemission spectral functions.

\section{Methodology}
The full Green's function for a single band system at 0 K can be
written as
\begin{eqnarray}
G^{Full}(\textbf{k},\omega)&=&\sum_n[\frac{|\langle
n|c_k|0\rangle|^2}{\omega-E_n+i\delta}+\frac{|\langle
n|c^\dag_k|0\rangle|^2}
{\omega-E_n+i\delta}]\nonumber\\
&=&\frac{1}{\omega-\varepsilon_k-\Sigma^{Full}(\textbf{k},\omega)}
\end{eqnarray}
where $\varepsilon_k$ and $\Sigma^{Full}$(\textbf{k},$\omega$) are
the bare band and self energy, respectively. The chemical
potential $\mu$ is set to zero for convenience. Note that due to
the sum rule $\sum_n\{|\langle n|c_k|0\rangle|^2+|\langle
n|c^\dag_k|0\rangle|^2\}$=1, the asymptotic behavior
($\omega\rightarrow\infty$) of $G^{Full}$ is 1/$\omega$. This
makes the self energy non-divergent when $\omega \rightarrow
\pm\infty$. In addition, the self energy is analytic in the
positive complex plane, and as a consequence the real and
imaginary parts of the self energy are Hilbert transforms of each
other\cite{Luttinger}. The Green's function defined in this way
(\textquotedblleft full Green's function\textquotedblright) is
thought to represent electron addition or removal processes and
the self energy holds the information on the interaction.

The full Green's function in the above equation however includes
not only electron removal but also electron addition process for
which we do not have information. Our goal is to obtain the self
energy from a general spectral function with as few assumptions as
possible. Therefore, we attempt to look at the self energy for the
electron removal Green's function which should best describe the
ARPES process and is defined as
\begin{eqnarray}
G^{ER}(\textbf{k},\omega)&=&\sum_n\frac{|\langle
n|c_k|0\rangle|^2}{\omega-E_n+i\delta}\nonumber\\
&=&\frac{n_k}{\omega-\varepsilon_k-\Sigma^{ER}(\textbf{k},\omega)}
\end{eqnarray}
or
\begin{eqnarray}
\Sigma^{ER}(\textbf{k},\omega)&=&\omega-\varepsilon_k-\frac{n_k}{G^{ER}(\textbf{k},\omega)}\nonumber\\
&=&\omega-\varepsilon_k-\frac{n_k}{\sum_n\frac{|\langle
n|c_k|0\rangle|^2}{\omega-E_n+i\delta}}
\end{eqnarray}
where $n_k=\sum_n|\langle n|c_k|0\rangle|^2=\int
A^{ER}(\textbf{k},\omega)d\omega$ and $A^{ER}(\textbf{k},\omega)$
is the electron removal spectral function (the superscript ER
stands for \textquotedblleft electron removal\textquotedblright).
Another difference is that $\varepsilon_k$ $may$ $not$ be the bare
band dispersion as will be discussed later. The asymptotic
behavior of $G$ is now $n_k/\omega$ and with the definition given
in equation (2), the real and imaginary parts of the electron
removal self energy (Re$\Sigma^{ER}$ and Im$\Sigma^{ER}$,
respectively) are again related through Hilbert transforms.

To obtain the electron removal self energy $\Sigma^{ER}$, we use
the fact that the measured ARPES spectrum is proportional to the
electron removal spectral function, that is,
\begin{eqnarray}
I_{ARPES}(\textbf{k},\omega)\equiv
I&=&|M(\textbf{k})|^2A^{ER}(\textbf{k},\omega)\nonumber\\
&=&-\frac{|M(\textbf{k})|^2}{\pi}\text{Im}G^{ER}(\textbf{k},\omega)
\end{eqnarray}
where \emph{I}$_{ARPES}$ and M(\textbf{k}) are the measured ARPES
spectrum and matrix element, respectively. Here, the $\omega$
dependence of the matrix element M(\textbf{k}) is ignored as
usual. Note that the Fermi function is missing as we deal with the
electron ${\it removal}$ Green's and spectral functions. The real
and imaginary parts of the electron removal Green's function are,
as is the case for $G^{Full}$, Hilbert transforms of each other,
Re$G^{ER}$=H(Im$G^{ER}$) where H() represents the Hilbert
transformation. Then the combination of equations (3) and (4)
gives\cite{Norman},
\begin{equation}
\text{Re}\Sigma^{ER}=\omega-\varepsilon_k-C(\textbf{k})\frac{H(I)}{H(I)^2+I^2}
\end{equation}
\begin{equation}
\text{Im}\Sigma^{ER}=C(\textbf{k})\frac{I}{H(I)^2+I^2}
\end{equation}
where $\varepsilon_k$ and $C(\bf k)$ are the bare band and
-$n_k|M(\bf k)|^2/\pi$, respectively.

The only remaining task to get $\Sigma$ is to determine $C(\bf k)$
and $\varepsilon_k$. We can obtain $C(\bf k)$ by integrating both
sides of equation (4).
\begin{eqnarray}
\int I_{ARPES}(\textbf{k},\omega)d\omega&=&-\frac{\pi
C(\textbf{k})}{n_k}\int
A^{ER}(\textbf{k},\omega)d\omega\nonumber\\
&=&-\frac{\pi C(\textbf{k})}{n_k}n_k=-\pi C(\textbf{k})
\end{eqnarray}
To define $\varepsilon_k$, we enforce the standard relationship
Re$\Sigma^{ER}$=H(Im$\Sigma^{ER}$). For this relationship to hold,
$\Sigma^{ER}$(\textbf{k},$\omega$) $\rightarrow$ 0 as $\omega
\rightarrow \pm\infty$. By expanding the last term in equation (3)
in terms of 1/$\omega$ for $\omega \rightarrow \pm\infty$, we get
\begin{eqnarray}
\Sigma^{ER}&=&\omega-\varepsilon_k-\frac{n_k\omega}{\sum_n\frac{|\langle
n|c_k|0\rangle|^2}{1-E_n/\omega}}\\
 &\approx& \omega-\varepsilon_k-\frac{n_k\omega}{\sum_n|\langle
n|c_k|0\rangle|^2+\sum_n|\langle
n|c_k|0\rangle|^2E_n/\omega}.\nonumber
\end{eqnarray}
Using $\sum_n|\langle n|c_k|0\rangle|^2=n_k$ and $\sum_n|\langle
n|c_k|0\rangle|^2E_n/n_k\equiv \bar{\varepsilon}_k$, $\Sigma^{ER}$
becomes
\begin{eqnarray}
\Sigma^{ER}=\omega-\varepsilon_k-\frac{n_k\omega}{n_k+n_k\bar{\varepsilon}_k/\omega}
\approx\bar{\varepsilon}_k-\varepsilon_k.
\end{eqnarray}
Since $\Sigma^{ER}$(\textbf{k},$\omega$) $\rightarrow$ 0 as
$\omega \rightarrow \pm\infty$, we set
$\varepsilon_k=\bar{\varepsilon}_k$ and thus we have
\begin{eqnarray}
\varepsilon_k=\frac{\int\omega
A^{ER}(\textbf{k},\omega)d\omega}{\int
A^{ER}(\textbf{k},\omega)d\omega}=\frac{\int\omega
I_{ARPES}(\textbf{k},\omega)d\omega}{\int
I_{ARPES}(\textbf{k},\omega)d\omega}.
\end{eqnarray}
That is, $\varepsilon_k$ is the center of gravity of the spectral
function but may not be the true bare band.

We rationalize this definition of $\varepsilon_k$ from the
knowledge that $\bar{\varepsilon}_k=\varepsilon_k$, the bare band
value, independent of the strength of interactions in the limit
$n_k \rightarrow$ 1\cite{Langreth}. Alternatively, $\varepsilon_k$
could have been chosen differently such as the true bare band
value of the original non-interacting problem but it will only
shift Re$\Sigma^{ER}$ determined here by a constant. In that case,
the relationship between Re$\Sigma^{ER}$ and Im$\Sigma^{ER}$ would
be modified to
Re$\Sigma^{ER}$=H(Im$\Sigma^{ER}$)+$\bar{\varepsilon}_k$-$\varepsilon_k$.
We note then that Re$\Sigma^{ER}$ goes to a finite value of
$\bar{\varepsilon}_k$-$\varepsilon_k$ as $\omega \rightarrow
\pm\infty$ (similarly Re$\Sigma^{Full}$ asymptotically goes to the
Hartree-Fock value). Using this expression we can see that the
parameter $\varepsilon_k$ drops out and that the imaginary part of
the self energy remains uniquely determined independent of our
choice of $\varepsilon_k$.

\begin{figure}
\centering \epsfxsize=8.5cm \epsfbox{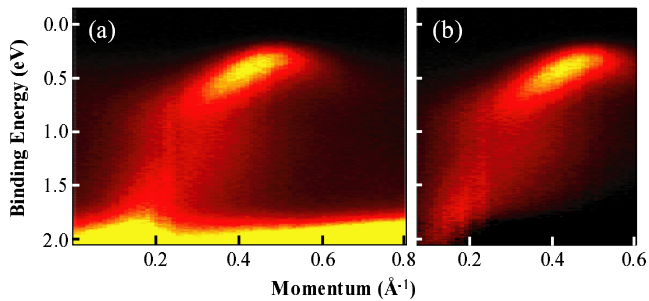} \caption{(Color
online) (a) Published ARPES data from CCOC along the (0,0) to
($\pi$,$\pi$) cut in Ref 8. (b) Same data after background is
removed. Data is presented for a smaller momentum window where
peaks are reasonably strong.}\label{fig1}
\end{figure}

\section{Results}
In applying the newly developed method to the ARPES data, one has
to consider other aspects of the experimental data such as
experimental resolution and thermal broadening. In that respect,
CCOC can be an ideal system to test the method: The experimental
resolution and thermal broadening are much smaller than the width
of the characteristic features of CCOC and thus can be ignored.
This provides an extra motivation for this study. For the
experimental data, we use the published CCOC data along the (0,0)
to ($\pi$,$\pi$) cut shown in Fig. 1(a)\cite{Ronning}.

Before we apply the procedure to extract $\Sigma^{ER}$, we also
have to subtract the background intensity from the data. This is
because the background intensity is from a process that can not be
described by the Green's function in equation (2). In our case,
subtracting the background so that $A^{ER}$(\textbf{k},$\omega$)
$\rightarrow$ 0 as $\omega \rightarrow \pm\infty$ is a necessary
condition for Re$\Sigma^{ER}$ and Im$\Sigma^{ER}$ to be Hilbert
transforms of each other. The background intensity is also
casually subtracted in other methods, either naturally by fitting
the MDCs\cite{McDougall} or enforcing it\cite{Norman}. There are a
number of potential ways to subtract the background, among which
we may consider the following two. The first one is subtracting
the Shirley background\cite{Shirley, Hufner}. The other is
subtracting the intensity of the unoccupied \textbf{k} states
(approximately beyond $k=0.7 {\AA}$ in Fig. 1(a)), that is, the
MDC background assuming that it is from the momentum independent
background intensity. We show the result extracted by using the
latter method in Fig. 1(b). However, the final results are similar
independent of the background subtraction method.

\begin{figure}
\centering \epsfxsize=8.5cm \epsfbox{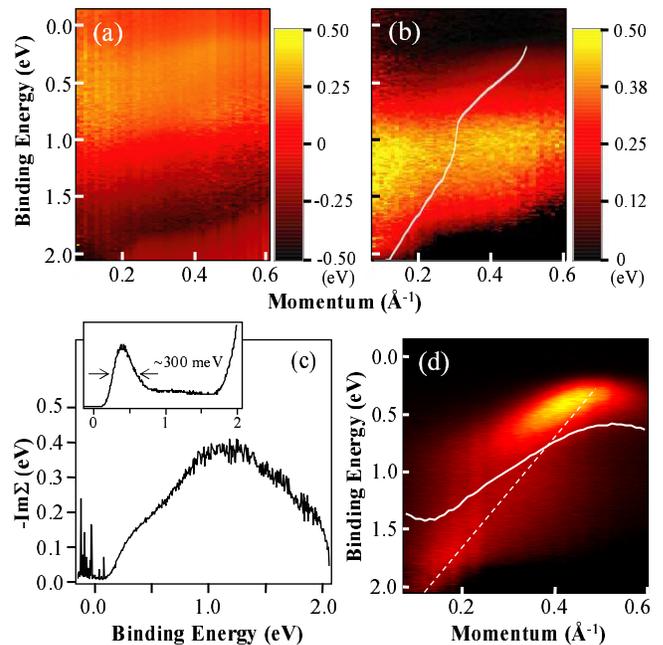} \caption{(Color
online) (a) Re$\Sigma^{ER}$ extracted from the data in Fig. 1(a).
The energy scale is coded in the color scales on the right. (b)
-Im$\Sigma^{ER}$. The line is the dispersion extracted from
fitting the MDCs. (c) -Im$\Sigma^{ER}$ along the MDC dispersion in
panel (b). The inset shows the EDC from the $k$ point where the
spectrum is the peakiest. The full width at half maximum of the
peak is about 300 meV. (d) Comparison of the $\varepsilon_k$
(solid line), and the MDC dispersion at the high binding energy
side and its extrapolation to the low energy side (dashed line).
ARPES data is shown in the background for the comparison purpose.
} \label{fig2}
\end{figure}

In panels (a) and (b) of Fig. 2, we show Re$\Sigma^{ER}$ and
Im$\Sigma^{ER}$ obtained by applying the procedure discussed
above. Note that while Re$\Sigma^{ER}$ can be negative,
-Im$\Sigma^{ER}$ is always positive. The most notable aspect of
the data is that overall $k$ dependence for both Re$\Sigma^{ER}$
and Im$\Sigma^{ER}$ is weak. This is very striking considering the
fact that the dispersion of the spectral features is very strong.
The momentum independence of the self energy which has been
casually
assumed\cite{McDougall,Khurana,Eschrig,Kordyuk_B,Kordyuk_L,Meevasana}
is indeed seen in the data. The momentum independence could
further be tested for our systems. Yet, we still see some momentum
dependence of the self energy, especially at the low energy side.
Therefore, for the accurate measure of the self energy, it is
important to have the self energy in the full \textbf{k}-$\omega$
space as is the case for our method.

The potential importance of the above observation can not be
understated. The largest shortcoming of dynamical mean field
theory (DMFT) is that momentum independence of the self energy
must be assumed (some momentum dependence can be added back in
with cluster calculations, but this is computationally very
expensive)\cite{A.Georges,G.Kotliar}. While experimental
limitations forced us to make a few assumptions such as the
background subtraction, our results are encouraging that the
momentum independent self energy assumption made in DMFT
calculations of the cuprates are valid. To our knowledge, this is
the first experimental attempt to prove the validity of this
assumption.

Looking at Im$\Sigma^{ER}$ plotted in panel (b), one may wonder
about the validity of the result near the gap region as
Im$\Sigma^{ER}$ is zero while it is known that Im$\Sigma^{Full}$
diverges near the gap region\cite{R.Bulla}. The divergence of
Im$\Sigma^{Full}$ near the gap is due to the fact that multiple
poles of $G^{Full}$ (or two peak structure in
$A^{Full}$(\textbf{k},$\omega$)) are attempted to be described by
a single pole with a self energy. However, by using only the
occupied part, \emph{A$^{ER}$} effectively becomes a single peak
spectral function. As Im$\Sigma^{Full}$ does not diverge for a
single peak spectral function, Im$\Sigma^{ER}$ calculated from an
effective single peak does not necessarily diverge even though
CCOC is an insulator. This issue will be discussed again with
simulations presented in Fig. 3.

In the conventional method of fitting MDCs and converting the MDC
widths to EDC widths\cite{McDougall}, one estimates the width,
thus Im$\Sigma^{ER}$, at the peak positions. We present the result
in a similar fashion by plotting Im$\Sigma^{ER}$ along the
dispersion. This provides the most accurate estimate of
$\Sigma^{ER}$ of the feature because of the possible momentum
dependence discussed above. In Fig. 2(c), we plot -Im$\Sigma^{ER}$
along the MDC dispersion shown in panel (b). At this point, it is
worth checking the validity of the new result. One way of doing so
is to compare the Im$\Sigma^{ER}$ obtained by the new method with
the value from a conventional method (if such a method can be
applied). In the case of CCOC, a conventional way of measuring the
half width of a quasiparticle-like peak may be applicable only to
the features with the lowest binding energy. In the inset, we plot
the EDC of the feature that has the lowest binding energy of 0.4
eV, hence is sharpest. The half width of the feature is about 150
meV. Im$\Sigma^{ER}$ at 0.4 eV indeed shows a similar value. This
fact gives us confidence in the new method.

Looking at Im$\Sigma^{ER}$ curve in panel (c), one sees
\textquotedblleft bends\textquotedblright at about 0.3 and 1 eV
binding energies. 0.3 eV is where the spectral weight of the
lowest binding energy feature starts as can be seen from Fig.
1(a). Therefore, the bend at 0.3 eV may not mean any coupling
energy. Meanwhile the bend at 1 eV is in the middle of the
dispersion. Interestingly, 1 eV binding energy is where the
kink-like dispersion appears to have \textquotedblleft bosonic
mode\textquotedblright coupling. When the kink-like feature from
CCOC was first reported\cite{Ronning}, the most obvious problem
with the interpretation of it in terms of bosonic mode coupling
was how an insulator can have bosonic mode coupling. Strictly
speaking, an insulator can have bosonic mode coupling and should
display kink-like features even though the effect would be
small\cite{Leem}. Even though it does not show the origin, our
analysis firmly shows that Im$\Sigma^{ER}$ behaves as if there is
a bosonic mode coupling at 1 eV (or 0.6 eV from the top of the
band). This value is comparable to the energy found in doped
HTSCs\cite{Xie,Graf}.

In Fig. 2(d), we plot the MDC dispersion of the high energy
features and its extrapolation as the dashed line. This has a
strong resemblance to the LDA band and may serve as the bare
band\cite{Ronning}. Also plotted in the panel is the
$\varepsilon_k$ obtained by calculating the center of the gravity
of the spectral function. In comparing the two, one sees very
little resemblance between them. The deviation of $\varepsilon_k$
from the bare band near $k=0.5\AA^{-1}$ (where $n_k\neq 1$) is
from the reason discussed above. Meanwhile, we believe the
difference near $k=0$, where we expect $n_k$ to be closer to 1, is
from the fact that we cut out the spectral weight at high binding
energies under the main valence band\cite{Eskes} during the
background subtraction.

\begin{figure}
\centering \epsfxsize=8.5cm \epsfbox{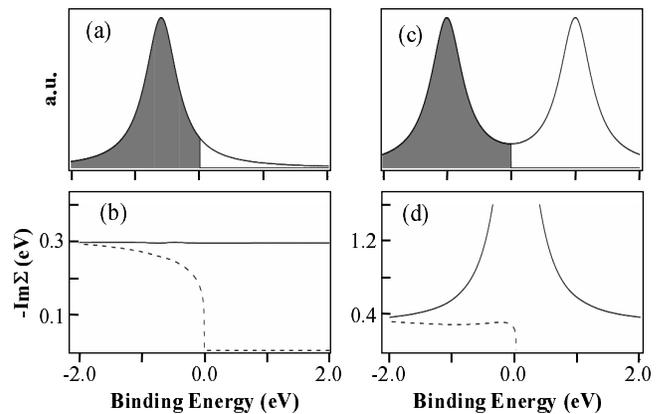} \caption{(a) A
Lorentzian peak with an width of 0.3 eV representing a spectral
function $A^{Full}(\textbf{k},\omega)$. The filled area below the
Fermi energy represents the electron removal spectral function
$A^{ER}(\textbf{k},\omega)$. (b) -Im$\Sigma^{Full}$ (solid) and
-Im$\Sigma^{ER}$ (dashed) of the spectral functions in (a). (c)
Two Lorentzian peaks of 0.3 eV width, simulating a situation where
there is a gap at the Fermi energy. The filled area again
represents $A^{ER}(\textbf{k},\omega)$. on the red solid line
(blue solid line). (d) -Im$\Sigma^{Full}$ (solid) and
-Im$\Sigma^{ER}$ (dashed) of the spectral functions in (c).}
\label{fig3}
\end{figure}

An important question is whether the self energy defined only for
the electron removal process is meaningful. We argue that it still
bares useful information. First of all, the self energy defined in
this way is identical to the usual $\Sigma^{Full}$ when $n_k=1$.
This is important because for most of the cases one analyzes the
self energy for the occupied parts of the momentum space where
$n_k\approx1$. In addition, we find that Im$\Sigma^{ER}$ measures
the \textquotedblleft width\textquotedblright of spectral function
fairly well in various cases as illustrated below. Lastly, on the
practical side, the method can also be applied when the dispersion
is flat and therefore MDCs can not be obtained. This is very
useful when the dispersion does not cross the Fermi energy as in
insulators. To discuss if Im$\Sigma^{ER}$ indeed measures the
width of the occupied part of a spectral function well, we use
numerical simulations as illustrated in Fig. 3. The first case,
shown in panel (a), is when the spectral weight is truncated by
the Fermi function. From the results plotted in panel (b), we find
that Im$\Sigma^{ER}$ still measures the width of the peak fairly
well below the Fermi energy. We may also discuss cases where there
is a gap at the Fermi energy and the spectral function splits into
two due to, for example, charge density wave or Mott transition.
To simulate such cases, we use two Lorentzian peaks as shown in
panel (c). From the self energies plotted in panel (d), we find
that Im$\Sigma^{ER}$ is very close to the Lorentzian peak width
while Im$\Sigma^{Full}$ has a large value near $\omega$=0. These
numerical simulation results support the assertion that
Im$\Sigma^{ER}$ measures the width of the occupied part of a
spectral feature. The case illustrated in panels (c) and (d) is
also relevant to the discussion on the behavior of the self
energies in the gap region. While Im$\Sigma^{Full}$ has a large
value at $\omega$=0, Im$\Sigma^{ER}$ stays at a finite value close
to the peak width.

Our method, as it is, may not be used for multi-band systems. This
is due to the fact that states from different bands may have
different matrix element $M(\textbf{k})$. However, our scheme in
principle can be extended to multi-band systems even though it is
more difficult. If it can be extended to multi-band system, our
method will find far more applicability and usefulness in the
analysis of self energies of many-body systems. This is at the
present left for the future work.

\section{Conclusion}
In conclusion, we propose using the electron removal self energy
which is found to represent the spectral width very well. The
method has the advantage that one can easily extract the self
energy from a spectral function with arbitrary shape, allowing us
to avoid making inappropriate assumptions. Application of the new
method to experimental data from CCOC reveals momentum
independence of $\Sigma$(\textbf{k},$\omega$) which has been an
assumption in the conventional analysis. We expect this new method
to be useful in strongly correlated systems where spectral
features are often quite broad.

\begin{acknowledgments}
Authors would like to thank B. J. Kim for a key suggestions in
programming and W. Meevasana, Jun Won Rhim and Jung Hoon Han for
helpful discussions. This work is supported by the KOSEF through
CSCMR and by the KICOS through a grant provided by MOST in
M60602000008-06E0200-00800. Work at Los Alamos was performed under
the auspices of US DOE.
\end{acknowledgments}


\begin{thebibliography}{22}

\bibitem[*]{Corres} Electronic address: cykim@phya.yonsei.ac.kr

\bibitem{McDougall} B. A. McDougall, T. Balasubramanian, and E. Jensen,
Phys. Rev. B {\bf51}, 13891 (1995).

\bibitem{Hengsberger} M. Hengsberger, D. Purdie, P. Segovia, M. Garnier, and Y. Baer, Phys. Rev. Lett.
{\bf 83}, 592 (1999).

\bibitem{Valla} T. Valla, A. V. Fedrov, P. D. Johnson, and S. L. Hulbert, Phys. Rev. Lett. {\bf83}, 2085 (1999)

\bibitem{Bogdanov} P.V. Bogdanov, A. Lanzara, S.A. Kellar, X.J. Zhou,
E.D. Lu, W.J. Zheng, G. Gu, J.-I. Shimoyama, K. Kishio, H. Ikeda,
R. Yoshizaki, Z. Hussain, and Z.-X. Shen, Phys. Rev. Lett.
{\bf85}, 2581 (2000).

\bibitem{Lanzara} A. Lanzara, P. V. Bogdanov, X. J. Zhoud, S. A. Kellar,
D. L. Feng, E. D. Lu, T. Yoshida, H. Eisaki, A. Fujimori, K.
Kishio, J.-I. Shinoyama, T. Noda, S. Uchida, Z. Hussain, and Z.-X.
Shen, Nature (London) {\bf 412}, 510 (2001).

\bibitem{Kaminski} A. Kaminski, M. Randeria, J. C. Campuzano,
M. R. Norman, H. Fretwell, J. Mesot, T. Sato, T. Takahashi, and K.
Kadowaki Phys. Rev. Lett. {\bf86}, 1070 (2001).

\bibitem{Gromko} A. D. Gromk, A. V. Fedorov, Y.-D. Chuang, J. D. Koralek,
Y. Aiura, Y. Yamaguchi, K. Oka, Yoichi Ando, and D. S. Dessau,
Phys. Rev. B {\bf 68}, 174520 (2003).

\bibitem{Meevasana} W. Meevasana, N. J. C. Ingle, D. H. Lu, J. R. Shi,
F. Baumberger, K. M. Shen, W. S. Lee, T. Cuk, H. Eisaki, T. P.
Devereaux, N. Nagaosa, J. Zaanen, and Z.-X. Shen, Phys. Rev. Lett.
{\bf 96}, 157003 (2006).

\bibitem{Ronning} F. Ronning, K. M. Shen, N. P. Armitage, A. Damascelli,
D. H. Lu, Z.-X. Shen, L. L. Miller, and C. Kim, Phys. Rev. B {\bf
71}, 094518 (2005).

\bibitem{Xie} B. P. Xie, K. Yang, D. W. Shen, J. F. Zhao, H. W. Ou, J. Wei,
S. Y. Gu, M. Arita, S. Qiao, H. Namatame, M. Tanaguchi, N. Kaneko,
H. Eisaki, Z. Q. Yang, and D. L. Feng, Phys. Rev. Lett. {\bf 98},
147001 (2007).

\bibitem{Graf} J. Graf, G.-H. Gweon, K. McElroy, S. Y. Zhou, C. Jozwiak,
E. Rotenberg, A. Bill, T. Sasagawa, H. Eisaki, S. Uchida, H.
Takagi, D.-H. Lee, and A. Lanzara, Phys. Rev. Lett. {\bf 98},
067004 (2007).

\bibitem{C.Kim} C. Kim, F. Ronning, A. Damascelli, D. L. Feng,
Z.-X. Shen, B. O. Wells, Y. J. Kim, R. J. Birgeneau, M. A.
Kastner, L. L. Miller, H. Eisaki and S. Uchida, Phys. Rev. B {\bf
65}, 174516 (2002).

\bibitem{K.M.Shen} K. M. Shen, F. Ronning, W. Meevasana, D. H. Lu,
N. J. C. Ingle, F. Baumberger, W. S. Lee, L. L. Miller, Y.
Kohsaka, M. Azuma, M. Takano, H. Takagi, and Z.-X. Shen, Phys.
Rev. B {\bf 75}, 075115 (2005).


\bibitem{B.J.Kim} B. J. Kim, H. Koh, E. Rotenberg, S.-J. Oh, H. Eisaki,
N. Motoyama, S. Uchida, T. Tohyama, S. Maekawa, Z.-X. Shen and C.
Kim, Nat. Phys. {\bf2}, 397 (2006).

\bibitem{Norman} M. R. Norman, H. Ding, H. Fretwell, M. Randeria, J. C. Campuzano,
Phys. Rev. B {\bf60}, 7585 (1999).

\bibitem{Hasan} M. Z. Hasan, E. D. Isaacs, Z.-X. Shen, L.L. Miller,
K. Tsutsui, T. Tohyama, and S. Maekawa, Science {\bf 288}, 1811
(2000).

\bibitem{Luttinger} J. M. Luttinger, Phys. Rev. {\bf121}, 942 (1961).

\bibitem{Langreth} D. C. Langreth, Phys. Rev. B {\bf1}, 471
(1970).

\bibitem{Shirley} D. A. Shirley, Phys. Rev. B {\bf5}, 4709 (1972).

\bibitem{Hufner} S. H\"{u}fner, {\it Photoelectron spectroscopy : principles and
application}, Oxford University, New York : Springer-Verlag, 1995.

\bibitem{R.Bulla} R. Bulla, T. A. Costi and D. Vollhardt, Phys. Rev. B {\bf 64}, 045103 (2001).

\bibitem{Khurana} Anil Khurana, Phys. Rev. B {\bf40}, 4316 (1989);
P. W. Anderson, Princeton University lecture notes on storonly
interacting fermions (unpublished).

\bibitem{Eschrig} M. Eschrig and M. R. Norman, Phys. Rev. B {\bf 67},
144503 (2003).

\bibitem{Kordyuk_B} A. A. Kordyuk, S. V. Borisenko, A. Koitzsch,
J. Fink, M. Knupfer, and H. Berger, Phys. Rev. B {\bf 71}, 214513
(2005).

\bibitem{Kordyuk_L} A. A. Kordyuk, S.V. Borisenko, V. B. Zabolotnyy, J. Geck,
M. Knupfer, J. Fink, B. Buchner, C. T. Lin, B. Keimer, H. Berger,
A.V. Pan, Seiki Komiya, and Yoichi Ando, Phys. Rev. Lett. {\bf
97}, 017002 (2006).

\bibitem{A.Georges} A. Georges, G. Kotliar, W. Krauth and M. J.
Rozenberg, Rev. Mod. Phys. {\bf68}, 13 (2006).

\bibitem{G.Kotliar} G. Kotliar, S. Y. Savrasov, K. Haule, V. S.
Oudovenko, O. Parcollet, C. A. Marianetti, Rev. Mod. Phys.
{\bf78}, 865 (2006).

\bibitem{Leem} C. S. Leem, B. J. Kim, Chul Kim, S. R. Park, T. Ohta,
E. Rotenberg, H.-D. Kim, M. K. Kim, H. J. Choi, and C. Kim,
unpublished.

\bibitem{Eskes}H. Eskes and R. Eder, Phys. Rev. B {\bf 54}, R14226 (1996).

\end{thebibliography}
\end{document}